\documentclass[pdflatex,sn-basic]{sn-jnl}

\usepackage{graphicx}%
\usepackage{amsmath,amssymb,amsfonts}%
\usepackage[title]{appendix}%
\usepackage{xcolor}%
\usepackage{textcomp}%
\usepackage{manyfoot}%
\usepackage{booktabs}%
\usepackage{bibunits}%

\makeatletter
\def\email#1{\global\advance\emailcnt by 1\relax%
\if@corauemail%
   \ifx\corrauthemail\@empty\gdef\corrauthemail{\setcounter{footnote}{0}\textcolor{blue}{#1}}%
   \else\g@addto@macro\corrauthemail{\ and\ \textcolor{blue}{#1}}\fi%
\else%
   \ifx\authemail\@empty\gdef\authemail{\setcounter{footnote}{0}\textcolor{blue}{#1}}%
   \else\g@addto@macro\authemail{\ and\ \textcolor{blue}{#1}}\fi%
\fi}
\makeatother

\makeatletter
\g@addto@macro\abstractfont{\unboldmath}
\makeatother

\begin{document}

\title[AI Disclosure Penalizes Source Authenticity More Than Intended Use]{AI Disclosure Penalizes Source Authenticity More Than Intended Use}

\author*[1]{\fnm{Minsu} \sur{Park}}\email{mp5500@nyu.edu}
\equalcont{These authors contributed equally to this work.}

\author[2]{\fnm{Hyunwoo} \sur{Kim}}
\equalcont{These authors contributed equally to this work.}

\author[2]{\fnm{Yoonseo} \sur{Choi}}

\author[3]{\fnm{Yoyo Tsung-Yu} \sur{Hou}}

\author[2]{\fnm{Juho} \sur{Kim}}

\affil[1]{\orgdiv{Division of Social Science}, \orgname{New York University Abu Dhabi}, \orgaddress{\city{Abu Dhabi}, \country{UAE}}}

\affil[2]{\orgdiv{School of Computing}, \orgname{KAIST}, \orgaddress{\city{Daejeon}, \country{Republic of Korea}}}

\affil[3]{\orgdiv{Digital Content and Technologies}, \orgname{National Chengchi University}, \orgaddress{\city{Taipei City}, \country{Taiwan}}}

\abstract{AI disclosure policies aim to make the use of AI transparent to users. Research on AI aversion suggests that people often judge AI-labeled content, and especially its author, more negatively on dimensions such as authenticity, but whether this makes them less willing to use the content is unclear. We tested this in two randomized online experiments on advice ($N = 1{,}732$). Labeling an answer as AI-generated rather than human-written substantially reduced perceived author authenticity, as in previous work, while perceived information quality and willingness to use the advice changed much less. Intended use was most closely associated with perceived information quality and least with perceived authenticity, so the evaluation most affected by disclosure was the one least associated with use. This pattern held across questions independently rated as more suitable for human or AI advice. These results suggest that negative reactions to AI disclosure do not necessarily lead people to reject AI-generated advice: disclosure mainly changed how people viewed who produced the advice, while intended use was more closely related to the advice's perceived quality.}


\maketitle

\section{Introduction}\label{sec:intro}

\noindent Disclosure is becoming a common policy response to AI-generated content; the EU's Artificial Intelligence Act, for example, requires certain AI-generated content to be labeled so that audiences can recognize AI involvement~\citep{euaiact2024}. Labels are meant to inform people, not discourage them from using AI content. But if people react negatively to AI authorship, as in algorithm aversion~\citep{dietvorst2015algorithm}, disclosure may lower trust even when the content itself is accurate and appropriate~\citep{wang2026ai}. Whether such reactions affect use is what we investigate here.

Experiments that label identical or comparable text as AI-generated typically find lower evaluations of the author or creator, including perceived authenticity and legitimacy~\citep{rae2024effects} and competence~\citep{bohm2023people}. Effects on evaluations of the message are less consistent: perceived effectiveness and accuracy fall in some studies~\citep{karinshak2023working,wang2026ai}, depend on prior attitudes toward AI in others~\citep{lim2024effect}, and are absent in others~\citep{gallegos2026labeling}. Downstream outcomes (e.g., intentions to share or policy support) tend to be affected less or not at all~\citep{bohm2023people,lim2024effect,wang2026ai,gallegos2026labeling}. These studies report perceptions and downstream outcomes side by side, but rarely ask whether penalized perceptions are associated with use.

We examine two points where the link between a label and use may break. First, a label may influence but not determine \textit{perceived} authorship, because people also infer who wrote the answer from the text itself using imperfect stylistic heuristics~\citep{jakesch2023heuristics,kovacs2024turing}. Second, the evaluation most affected by disclosure may not be the one most closely associated with intended use, given that advice-taking research separates evaluation of an advisor from evaluation of the advice~\citep{bonaccio2006advice,bailey2023meta}. If the label mainly changes how people see the author, people may react more strongly to the source than to the information itself, and, in turn, their willingness to use the information may change much less. We test the first point by measuring the alignment between the disclosed label and participants' perceived authorship, and the second by comparing how strongly each evaluation responds to the randomized label with how closely it is associated with intended use.

We conducted two between-participant experiments testing how authorship labels affect evaluations and intended use of answers to advice-seeking questions, chosen from an independent survey to span topics people considered better suited to human advice, to either, or to AI. We use this question-and-answer setting because an answer's author and its information can come apart, and intended use can be operationalized as behavioral intentions (e.g., endorsing, sharing, or following the advice). US adults recruited from Prolific (Study 1, $N=839$; Study 2, $N=893$) read one question with a single answer, wrote a brief summary, and evaluated author authenticity, information quality, personal resonance, readability, and intended use (details in Materials and Methods and SI Appendix, Sections~\ref{si:concept} and~\ref{si:measures}). Study 1 used six questions with a human, AI, or no authorship label; Study 2 reused three with a human or AI label. All answers were generated or rewritten with GPT-4o and calibrated so authorship was not obvious from style; the same answer appeared across label conditions to isolate the effect of presented authorship.

\section{Results}\label{sec:results}

\begin{figure}[t]
    \centering
    \includegraphics[width=\textwidth]{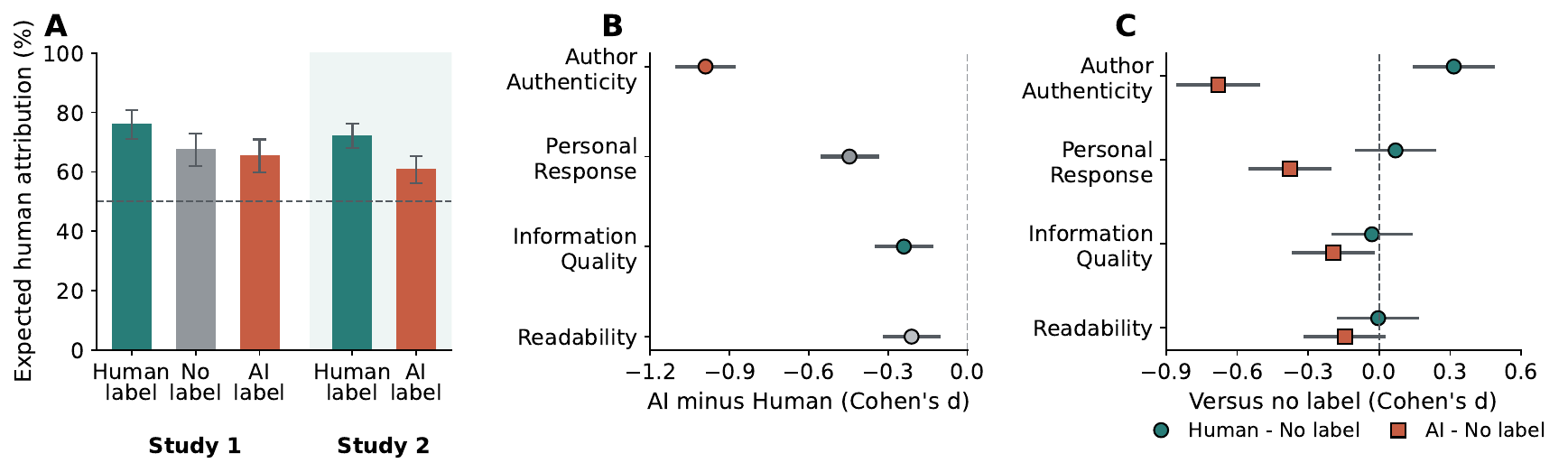}
    \caption{\textbf{Authorship disclosure selectively reshapes source evaluation.} (\textbf{A}) Percentage expecting other readers to attribute the answer to a human, by label condition (in Study 1, 76.4\%, 67.6\%, and 65.6\%, respectively, for human, undisclosed, and AI; in Study 2, 72.3\% and 61.0\%, respectively, for human and AI). (\textbf{B}) Fixed-effect AI-minus-human effects, synthesized across studies: author authenticity $d=-0.99$; personal resonance $-0.45$; information quality $-0.24$; readability $-0.21$. (\textbf{C}) Study 1 effects relative to no label: human label raised authenticity ($d=0.32$) only; AI label reduced authenticity ($-0.68$), personal resonance ($-0.38$), and information quality ($-0.19$).
    }
\label{fig:identity}
\end{figure}

\subsection{Labels shift but do not determine expected attribution}
After completing the evaluations, participants reported who other readers would believe had written the answer. We used this indirect measure of expected attribution to reduce the tendency to simply repeat the assigned label~\citep{fisher1993social}. In Study 1, 76.4\% expected a human author under the human label, 65.6\% under the AI label, and 67.6\% when undisclosed; in Study 2, 72.3\% and 61.0\% for the human and AI labels (Fig.~\ref{fig:identity}A). Relative to no label, a human label increased expected human attribution by 8.8 percentage points, whereas an AI label decreased it by only 2.0. Most participants expected other readers to perceive a human author even when the answer was labeled AI-generated. Disclosure is thus an informational cue, not a guarantee that inferences follow it, consistent with evidence that people judge authorship from text using unreliable heuristics~\citep{jakesch2023heuristics,kovacs2024turing}.

\subsection{Disclosure produces an authenticity penalty}
The label's penalty was concentrated on the author (Fig.~\ref{fig:identity}B). The combined AI-minus-human effect across studies on author authenticity was $d=-0.99$ [$-1.10$, $-0.88$], replicating in Study 1 ($d=-0.98$) and Study 2 ($d=-1.00$). Effects were much smaller for personal resonance ($d=-0.45$ [$-0.55$, $-0.34$]), information quality ($-0.24$ [$-0.34$, $-0.14$]), and readability ($-0.21$ [$-0.31$, $-0.11$]). In a joint model of all four evaluations, with standard errors clustered by participant, the authenticity penalty was significantly larger than each of the others (all $P<0.001$), indicating that disclosure mainly changed how participants judged the apparent author.

The undisclosed condition in Study 1 separated a human-label premium from an AI-label penalty (Fig.~\ref{fig:identity}C). Relative to no label, a human label raised authenticity ($d=0.32$ [$0.15$, $0.48$]) but did not reliably change any other evaluation, so the premium appeared confined to the author, not the answer's clarity, reasoning, or usefulness. An AI label lowered authenticity far more ($d=-0.68$ [$-0.85$, $-0.51$]), with smaller reductions in personal resonance and information quality. Note that the no-label baseline is not neutral because most participants in the undisclosed condition already expected a human source. These contrasts describe disclosure relative to ambiguity, not a clean split between human favoritism and AI aversion.

\subsection{Intended use declines less and is more closely associated with information quality}
\begin{figure}[t]
    \centering
    \includegraphics[width=0.7\textwidth]{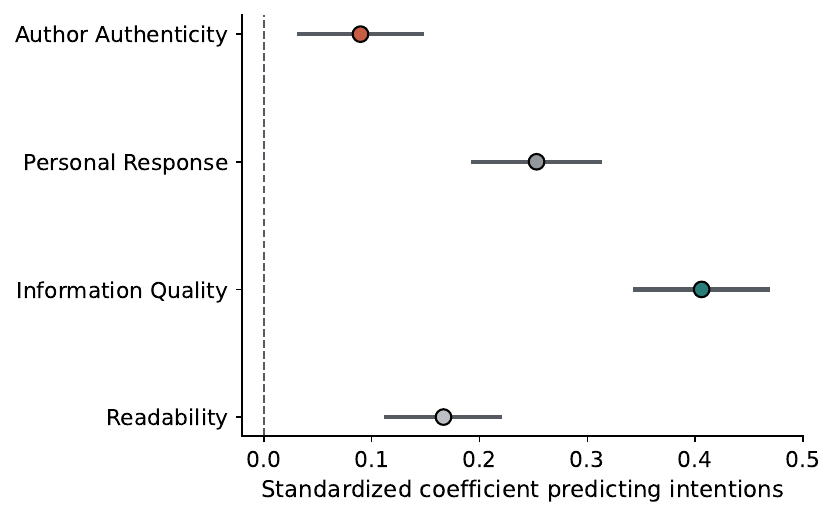}
    \caption{\textbf{What predicts intended use.} Standardized coefficients with 95\% confidence intervals from a joint model of intended use including all four evaluations, adjusted for label, study, and question: $0.090$ for author authenticity, $0.253$ for personal resonance, $0.406$ for information quality, $0.167$ for readability. The information quality coefficient exceeded the authenticity coefficient by $0.316$ ($P<0.001$).}
\label{fig:inversion}
\end{figure}

The AI label also reduced intended use ($d=-0.44$ [$-0.54$, $-0.33$]), i.e., behavioral aversion was attenuated, not absent. Importantly, the evaluations most affected by disclosure were not the one most closely related to intended use (Fig.~\ref{fig:inversion}). In a standardized model predicting intended use, information quality had the largest conditional association ($\beta=0.406$ [$0.344$, $0.468$]), while author authenticity had the smallest ($\beta=0.090$ [$0.032$, $0.147$]), controlling for label, study, question, and the other evaluations. A Shapley decomposition confirmed this result: information quality accounted for 37.5\% of incremental $R^2$, compared with 15.2\% for authenticity. Because the evaluations were measured rather than randomized, these are associations, not causal effects. Still, we found the same pattern across tasks seen as more suitable for a human or an AI, a dimension known to affect reliance on algorithms~\citep{castelo2019task,longoni2019resistance}; Bayesian models favored no moderation across outcomes (SI Appendix, Section~\ref{si:bayes}).

\section{Discussion}\label{sec:discussion}

AI disclosure had its largest effect on author authenticity. Intended use, however, was much more closely related to information quality. The evaluation most affected by disclosure was therefore the one least associated with intended use. The AI label still reduced intended use, so aversion persisted, although much less than authenticity. This extends prior work reporting that AI labels lowered evaluations of the creator but not of the content~\citep{bohm2023people,rae2024effects} by relating the penalty to evaluations that predict use. It may also explain why labels leave downstream outcomes little changed~\citep{gallegos2026labeling}, since their effect falls mainly on judgments of the author, which are less closely tied to use.

This distinction is also relevant for policy. A large penalty on source perceptions need not imply a comparable reduction in use, as the label had a much larger impact on authenticity compared to its impact on intended use. Conversely, continued use does not mean the label had no effect, because users may penalize an AI source while still using content they find instrumentally useful. Disclosure, recognition of AI involvement, source perceptions, and use should therefore be evaluated separately, not conflated.

This work has several limitations. First, intended use was self-reported, and although labels were randomized, evaluations were measured, precluding causal conclusions about which perceptions drive use. Second, all answers were AI-generated and labeled ``knowledgeable human contributor'' or ``knowledgeable AI,'' without other cues that may be visible in real-world settings (e.g., institutional affiliation or platform), so findings concern presented authorship in decontextualized conditions and may not extend to skeptical AI labels. Third, we used six advice questions and US participants; robustness across questions spanning the human-AI suitability spectrum is encouraging, but generalization needs further work.

These findings are a snapshot of a fast-moving target. As models improve, answers may become more accurate and harder to distinguish from human writing, making information quality an even stronger guide to use than source impressions. As AI becomes more familiar, the label may signal efficiency or accuracy rather than inauthenticity, so the author penalty could shrink, persist, or reverse. Because label effects likely depend on model quality, advice type, and how systems are deployed and regulated, future studies could explore whether and how these reactions change.

\section{Materials and Methods}\label{sec:methods}

\noindent \textbf{Participants.} We recruited US English-speaking Prolific participants with approval rates above 90\%. Studies ran in January and September--October 2025, respectively. Quality-control rules invalidated 38 Study 1 and 67 Study 2 records (SI Appendix, Section~\ref{si:participants}), leaving the samples reported above. Participants consented and received GBP 1.50 for a survey averaging seven minutes. The KAIST IRB approved the research (KH2024-049). Neither study was preregistered.

\noindent \textbf{Materials.} We ranked 173 advice-seeking questions in pairwise comparisons by whether respondents would direct them to a person or an AI; upper, middle, and lower thirds divided questions into human-preferred, neutral, and AI-preferred categories. Six questions matched to Reddit questions were selected across these categories. Answers were generated or rewritten with GPT-4o and revised until perceived authorship was approximately balanced, then fixed within question across label conditions.

\noindent \textbf{Measures.} Participants rated each construct on multi-item seven-point scales: intended use (endorse, share, follow), author authenticity (qualification, originality, sincere effort), information quality (sufficiency, objectivity, logic, factual correctness), personal resonance (emotional support, trust, surprise), and readability (clarity, concision). Items were averaged into participant-level composites. Expected attribution was elicited indirectly, asking who other readers would believe wrote the answer, to reduce social desirability bias~\citep{fisher1993social} (SI Appendix, \textit{Measures and composite scoring}).

\noindent \textbf{Analysis.} We report standardized mean differences with 95\% confidence intervals. Label effects were compared across dimensions in a regression controlling for study and question, with participant-clustered standard errors; separate per-study models confirmed robustness. The intended-use model used the four standardized evaluations simultaneously, controlling for label, study, and question; Shapley decomposition assessed relative contribution. Bayesian moderation models coded suitability as AI-preferred ($-1$), neutral ($0$), or human-preferred ($+1$). See SI Appendix for full details.

\backmatter

\bmhead{Supplementary information}
The SI Appendix follows the references.

\bibliography{aidisclosure}

\clearpage
\setcounter{section}{0}\setcounter{figure}{0}\setcounter{table}{0}\setcounter{equation}{0}
\renewcommand{\thesection}{S\arabic{section}}\renewcommand{\theHsection}{S\arabic{section}}
\renewcommand{\thefigure}{S\arabic{figure}}\renewcommand{\theHfigure}{S\arabic{figure}}
\renewcommand{\thetable}{S\arabic{table}}\renewcommand{\theHtable}{S\arabic{table}}
\renewcommand{\theequation}{S\arabic{equation}}\renewcommand{\theHequation}{S\arabic{equation}}

\begin{center}
{\Titlefont Supplementary Information for\\
AI Disclosure Penalizes Source Authenticity More\\
Than Intended Use\par}
\vspace{12pt}
{\large Minsu Park, Hyunwoo Kim, Yoonseo Choi,\\ Yoyo Tsung-Yu Hou, and Juho Kim}\\[6pt]
Corresponding author: \href{mailto:mp5500@nyu.edu}{mp5500@nyu.edu}
\end{center}
\bigskip

\begin{bibunit}[sn-basic]
\makeatletter\def\@extra@b@citeb{SI}\def\@extra@binfo{SI}\makeatother
\section{Conceptual basis for the evaluation dimensions}\label{si:concept}
People do not necessarily judge online information as a whole. Research on online credibility has shown that people tend to evaluate the source, the message, and their own response to it separately~\citep{wathen2002credibility}. Research on trust in AI makes a similar distinction between cognitive and affective aspects of trust~\citep{glikson2020trust}. Advice-taking research also separates evaluation of the advisor from evaluation of the advice~\citep{bonaccio2006advice}. We expected a similar distinction here. Since an authorship label directly tells participants who wrote the answer, we expected a significant change in how they viewed the author. This expectation is consistent with prior disclosure studies that show AI attribution can devalue the perceived creator or advisor even when content evaluations change less~\citep{bohm2023people,rae2024effects}. We therefore treated author authenticity as the focal source-directed dimension, distinguished it from three message- and reader-directed dimensions, and measured intended use as a separate downstream outcome. Prior studies have compared evaluative and downstream outcomes and found that label effects can differ across them~\citep{bohm2023people,lim2024effect,wang2026ai}, but these outcomes were generally analyzed as parallel endpoints rather than to test whether the evaluations most responsive to disclosure were also those most closely associated with use. We investigate this link explicitly in the present study.

Accordingly, we used five dimensions in the study: (a) author authenticity captures whether the apparent author seems qualified, original, and sincere; (b) information quality captures whether the answer seems sufficient, objective, logical, and factually correct; (c) personal resonance captures the respondent's affective and interpersonal response to the answer; (d) readability captures its clarity and concision; and (e) intended use separately measures willingness to endorse, share, and follow the advice (see the section on Measures and composite scoring for more details). A confirmatory factor analysis supported this five-factor structure over a one-factor model (five-factor: CFI \(=0.949\) and \(0.957\), RMSEA \(=0.077\) and \(0.074\); one-factor: CFI \(=0.833\) and \(0.815\), RMSEA \(=0.131\) and \(0.146\), for Studies 1 and 2, respectively).

\section{Study overview and estimand}\label{si:overview}
Studies 1 and 2 were between-participant online experiments. Each participant evaluated one answer to an advice-seeking question, randomly drawn from questions that varied in whether they were seen as more suitable for human or AI advice. In Study 1, participants were randomly assigned to a human, AI, or undisclosed-source label, while Study 2 randomly assigned participants to either a human or AI label. The answer itself was otherwise held constant across conditions. This design isolates the effect of presented authorship, rather than comparing advice actually written by humans and AI.

We also asked participants who they thought other readers would say had written the response. We use this as our measure of \emph{expected attribution}. We did not ask participants directly who they themselves thought wrote it because they had just seen a human or AI label. A direct question at that point could make the manipulation obvious or simply encourage them to repeat the label. Asking about other readers was intended to reduce such possible bias. We asked the question after all five main outcome measures, so it could not affect those evaluations. Expected attribution was measured rather than randomized. Label effects on it are randomized comparisons, but analyses relating it to other measures are descriptive rather than causal.

\section{Participants, recruitment, and ethics}\label{si:participants}
We recruited U.S. adults through Prolific. Participants had to be fluent in English and have a Prolific approval rate above 90\%. Before starting the study, they reviewed an online consent form. Participants were paid GBP 1.50 for completing the study, which takes 7 minutes on average. Study 1 was conducted in January 2025, and Study 2 in September and October 2025. The Study 1 and Study 2 datasets contained 877 and 960 records, respectively. We excluded records marked as invalid in the project data, leaving 839 participants in Study 1 and 893 in Study 2. All participants included in the final samples had complete responses for the items used to construct the five main outcome measures. The Institutional Review Board at KAIST approved the research (KH2024-049). Neither study was preregistered.

\section{Task suitability and stimulus selection}\label{si:suitability}
If the effect of disclosure depends on the type of task, results based on a single domain may partly reflect that domain rather than AI disclosure more generally. We therefore sampled advice questions that differed in how suitable they seemed for human or AI advice, rather than using questions from a single domain, and tested whether the results were consistent across them. In particular, we examined whether the label effects and the ordering of predictors of intended use reported in the main text were similar across this range. The Bayesian moderation analyses below provide this test.

We built this range of question categories using a collaborative pairwise WikiSurvey~\citep{salganik2015wiki}. The survey began with 10 advice-seeking questions written by the researchers. Participants saw pairs of questions and chose which one they would be more likely to ask another person on the Internet rather than an AI. They could also add new questions, which were then shown to later participants.

The candidate pool contained 193 questions. Two researchers reviewed them and removed 20 that were duplicates, invalid, or outside the intended scope of advice seeking, leaving 173 questions. Each question received a WikiSurvey score estimating how likely it was to be chosen over a randomly selected alternative. We divided the questions into three groups based on these scores: the upper third was classified as human-preferred, the middle third as neutral, and the lower third as AI-preferred. We then used affinity diagramming to group the questions by topic and selected two from each group. For each selected question, we found a closely matching Reddit post so that the final stimuli retained the detail and context of naturally occurring online advice seeking.

Study 1 used all six questions: relationship and career questions from the human-preferred group, finance and productivity from the neutral group, and gardening and exercise from the AI-preferred group. Study 2 reused three of them, one from each group: relationship, finance, and gardening. Importantly, all six questions were developed together and used in Study 1; Study 2 later reused three of them.

These categories reflect the relative rankings from the separate stimulus-development study. They do not mean that a topic is inherently more suitable for a human or an AI. The moderation analyses therefore ask whether the results are similar across the range of tasks we sampled, rather than whether they would be the same for every possible task or domain.

\section{Answer construction}\label{si:answers}
We wanted the answers to be informative and read naturally, without making it obvious whether they had originally been written by a person or generated by AI. For the three topics later reused in Study 2 (relationship, finance, and gardening), we prepared two candidate questions per topic. Each question had two answer variants: a human-written Reddit response rewritten with GPT-4o and a GPT-4o-generated response rewritten or humanized using the same model. The rewriting prompt asked GPT-4o to preserve the main ideas, level of detail, tone, and emotional content, keeping the answers around 200 words. We then conducted a blind survey with 80 participants, who judged whether the answers were written by a human or AI. Each participant evaluated one randomly assigned question-and-variant combination for each topic, with the topic order randomized. This gave us approximately 20 judgments per combination. We selected the combination closest to an even split between human and AI judgments for each topic. The final relationship and finance answers originally came from Reddit, while the gardening answer was generated by GPT-4o. The percentages judged as human-written were 60.0\% (12/20), 57.1\% (12/21), and 55.0\% (11/20), respectively. None of them differed statistically significantly from 50\% (exact binomial \(P>0.5\) for each), although the small number of raters meant that we could not rule out a moderate preference in either direction.

For the remaining three topics (career, productivity, and exercise), we used GPT-4o to generate answers with a prompt adapted from \citet{zhang2024coauthor}. The prompt asked for a sincere, neutral answer in the style of an online forum, no longer than 200 words. It also asked the model to avoid lists, greetings, exclamation marks, and emoji, while allowing first-person language, colloquial expressions, and minor imperfections. We reviewed and edited these answers before including them in the experiments. Unlike the other three topics, these answers were not evaluated in the blind survey. Of the six final answers, four originated from GPT-4o and two from Reddit, although all six were generated or rewritten with GPT-4o.

The selected answers were somewhat more likely to be judged human-written in the pilot. We found a similar pattern in the main experiments, where most participants in the no-label condition expected other readers to perceive a human author (Fig.~\ref{fig:identity}A). However, the two measures are not directly comparable. The pilot asked participants who they thought had written the answer, whereas the main experiments asked who they thought other readers would believe had written it. We therefore treat the no-label condition as a human-leaning baseline rather than a neutral one. Once selected, the final answers remained identical across the experimental label conditions. The stimulus archive includes the final questions and answers, URLs for the source questions, and the available prompts.

\section{Study 1 design and procedure}\label{si:study1}
In Study 1, participants were randomly assigned to one of six questions and one of three authorship conditions: Human, AI, or undisclosed source. In the Human condition, the answer appeared under the heading ``Answer by another person with relevant experience,'' while the AI condition used ``Answer by a knowledgeable artificial intelligence bot.'' When authorship was undisclosed, the heading simply read ``Answer,'' without identifying the source.

Participants first rated how suitable the question was for a human or AI respondent. They then read the answer and wrote brief summaries of both the question and the answer in their own words. After that, they completed the main evaluation and intended-use items. Later in the questionnaire, we asked about expected attribution, familiarity with and experience of the topic, familiarity with AI and attitudes toward it, and demographic characteristics. The expected-attribution question was the same for everyone, including participants who had not been shown an authorship label.

\section{Study 2 design and procedure}\label{si:study2}
Study 2 used three of the questions from Study 1. In this study, participants were assigned either a Human or an AI authorship label, along with one of three social-signal conditions: no additional signal, author reputation, or peer endorsement. In the reputation condition, participants were told that the author consistently ranked among the top 5\% of contributors for answer quality. In the endorsement condition, they were told that many people had found the particular answer helpful, and the response displayed 1,368 likes. Participants were randomly assigned to a question, an authorship label, and a social-signal condition. We followed the same questionnaire sequence as in Study 1, from reading the answer and writing a summary to completing the main measures, attribution question, and background questions.

The answer headers in Study 2 were ``Answer by a knowledgeable human contributor'' and ``Answer by a knowledgeable AI.'' Although these differed from those used in Study 1, both studies described the source as competent. We also found nearly identical effects on author authenticity across the two studies (Study 1, \(d=-0.98\); Study 2, \(d=-1.00\)). The social signals did not have statistically significant effects on any outcome, so we focus on the label effects common to both studies in the main text. We nevertheless included the social-signal conditions as controls in the intended-use and Bayesian moderation models. Their estimates can also be reproduced from the deposited data and code.

\section{Measures and composite scoring}\label{si:measures}
All main measures used seven-point response scales. Intended use ranged from \emph{extremely unlikely} (1) to \emph{extremely likely} (7), while the evaluation items ranged from \emph{strongly disagree} (1) to \emph{strongly agree} (7). Higher scores indicated more of the corresponding construct. We calculated each composite as the unweighted mean of its items. No imputation was needed for these items in the retained samples.

\begin{itemize}

\item[-] \textbf{Intended use} measured willingness to endorse or upvote an answer, share it with someone asking a similar question, and follow the advice.

\item[-] \textbf{Author authenticity} measured whether the author seemed qualified and trustworthy, whether the answer appeared original rather than copied, and whether the author seemed to have made a sincere effort.

\item[-] \textbf{Information quality} measured whether the answer seemed factually correct, sufficient to address the question, rational and logically sound, and objective and impartial.

\item[-] \textbf{Personal resonance} measured whether the answer provided appropriate emotional support, inspired belief or trust, and offered ideas that were surprising or intriguing.

\item[-] \textbf{Readability} measured whether the answer was concise and easy to read and understand.

\end{itemize}

To measure expected attribution, we asked, ``If other people read this same answer, who do you think they would believe wrote it?'' Participants chose between another person with relevant experience and an AI. We asked about other readers rather than directly asking participants who they thought the author was, to reduce social desirability bias and the pressure to repeat the label they had just seen or give the answer they believed the experimenter wanted~\citep{fisher1993social}. This wording also has a limitation: it captures what participants expected other readers to believe, not necessarily their own private belief. We used the same question in every condition, including the undisclosed condition in Study 1, allowing us to compare expected attribution across conditions.

\section{Parallel indirect-effects decomposition}\label{si:indirect}
We also estimated parallel indirect effects among participants assigned either a human or an AI label ($N=1{,}460$). The model estimated the association between the randomized label and each of the four evaluations, as well as the conditional association between each evaluation and intended use when all four were included together. Confidence intervals were calculated using 5{,}000 participant-level bootstrap samples.

The combined indirect association through the four evaluations was $-0.437$ (bootstrap 95\% CI, $-0.565$ to $-0.314$), accounting for 73.5\% of the total label difference in intended use. The label had a much larger effect on author authenticity than on the other evaluations, but its indirect association with intended use was relatively small ($ab=-0.104$, bootstrap 95\% CI, $-0.177$ to $-0.034$). For information quality, the indirect association was $-0.138$ ($-0.205$ to $-0.075$). The two indirect effects were not significantly different, as the larger label effect on authenticity compensated for its weaker association with use.

The indirect-effects results are consistent with the pattern reported in the main text, but they do not establish causal mediation. Unlike the authorship label, the four evaluations were measured rather than randomized. Whether changing answer quality would actually change use remains to be tested with an independent manipulation of quality and a consequential measure of information use.

\section{Bayesian task-suitability moderation}\label{si:bayes}
Task suitability was coded as AI-preferred (\(-1\)), neutral (0), or Human-preferred (\(+1\)). We restricted the moderation analyses to the Human- and AI-label conditions, with outcomes standardized within study. The label-effect models included the AI-label indicator, study, the Study 2 social-signal indicators, question controls, and interactions of the label with study and task suitability. We also fit a second model with all four evaluation predictors and their interactions with suitability. Here, we wanted to know whether the difference between the information-quality and author-authenticity coefficients changed across the tasks we sampled.

We used a conjugate Bayesian linear model with zero-centered normal priors conditional on residual variance. Prior variance was set to 10 for the intercept, 1 for ordinary coefficients, and \(0.50^2\) for moderation coefficients. For residual variance, we used an inverse-gamma prior with shape 2 and scale 2, and the coefficient summaries were based on the resulting Student-\(t\) posterior. The analysis archive reports posterior means, 95\% credible intervals, posterior mass within \(\pm0.05\) and \(\pm0.10\), and Savage--Dickey density-ratio Bayes factors for the point-null interactions. As a sensitivity check, we repeated the moderation analyses using prior standard deviations of 0.25, 0.50, and 1.00 for the moderation coefficients.

\section{Data preparation and privacy}\label{si:privacy}
Before creating the public analysis files, we removed direct identifiers and other information that could potentially identify participants, including IP addresses, Prolific IDs, Qualtrics response IDs, geographic coordinates, timestamps, and open-text responses. We also assigned new sequential participant IDs separately for each study. The released data retain randomized assignments, stimulus IDs, perceived-authorship measures, focal item responses, and calculated composites. The final datasets are provided as separate CSV files for each study and as a combined file containing one row per participant.

\section{Generative AI use statement}\label{si:ai}
Generative AI was used in three ways. (1) Research materials: OpenAI's GPT-4o was used to generate or rewrite the advice answers shown to participants, as described in Materials and Methods and the SI Appendix. (2) Manuscript preparation: Anthropic's Claude (Claude Sonnet 5.5) was used to suggest edits to the wording and length of the manuscript, SI Appendix, and cover letter; to check consistency among the text, SI Appendix, and figure; to retrieve and summarize cited literature for the authors to verify; and to compute descriptive statistics for the stimulus pilot (exact binomial tests and Wilson confidence intervals). (3) Replication package: GPT-5.6 Sol was used to clean up the released code and data files, organize them into a structure specified by the authors, and draft the README. M.P. and H.K. wrote the analysis code and conducted all analyses before this step. The authors designed the study, made all scientific and editorial decisions, checked every statement against the cited sources and the data, and take full responsibility for the content. No AI tool is listed as an author, and no AI-generated images or graphics were used.

\putbib[aidisclosure]
\end{bibunit}

\end{document}